\documentclass[amsmath, amssymb, aps, twocolumn]{revtex4-1}

\usepackage{graphicx}
\usepackage{amsmath}
\usepackage{bm}
\usepackage{booktabs}

\begin{document}

\title{Crystallographic and superconducting properties of the fully-gapped noncentrosymmetric 5$d$-electron superconductors Ca$M$Si$_3$ ($M=$ Ir, Pt)}

\author{G. Eguchi}
\email{geguchi@scphys.kyoto-u.ac.jp}
\author{D. C. Peets} \author{M. Kriener} \author{Y. Maeno}
\affiliation{Department of Physics, Graduate School of Science, Kyoto University, Kyoto 606-8502, Japan}

\author{E. Nishibori} \author{Y. Kumazawa} \author{K. Banno} \author{S. Maki} \author{H. Sawa}
\affiliation{Department of Applied Physics, Graduate School of Engineering, Nagoya University, Nagoya 464-8603, Japan}

\date{\today}

\begin{abstract}
We report crystallographic, specific heat, transport, and magnetic properties of the recently discovered noncentrosymmetric 5\textit{d}-electron superconductors CaIrSi$_3$ ($T_{\rm{c}}=3.6$~K) and CaPtSi$_3$ ($T_{\rm{c}}=2.3$~K). The specific heat suggests that these superconductors are fully gapped. The upper critical fields are less than 1~T, consistent with limitation by conventional orbital depairing. High, non-Pauli-limited $\mu_0H_{\rm c2}$ values, often taken as a key signature of novel noncentrosymmetric physics, are not observed in these materials because the high carrier masses required to suppress orbital depairing and reveal the violated Pauli limit are not present.
 
\begin{description}
\item[PACS numbers]
61.66.Fn, 74.20.Rp, 74.25.Dw
\end{description}
\end{abstract}

\pacs{Valid PACS appear here}

\maketitle

\section{\label{sec:level1}Introduction}


There has been a great deal of interest in noncentrosymmetric superconductors initiated by the discovery of highly unconventional behavior in CePt$_3$Si~\cite{Bauer2004PRL}. The overwhelming majority of superconductors studied to date have crystal structures exhibiting inversion symmetry.  If an inversion element is present, inversion about the origin in momentum space can at most change the sign of the superconducting pairing function, which can thus be classified by its parity, and the spin state of the pairs must be either symmetric or antisymmetric (triplet or singlet) on exchange of the component fermions.  In structures lacking an inversion element, however, there is no such constraint, parity is not a meaningful label, and pairing states cannot be classified as singlet or triplet.
In such a material,
spin-orbit terms can split the underlying band structure, and thus the Fermi surface, by spin orientation.
Where the splitting is large compared to the superconducting gap, pairing is expected to occur only within each Fermi surface sheet, and since there is only one allowed spin orientation at any given point on each sheet, the pairing state will be a mixture of what would normally be regarded as singlet and triplet components.  A wide variety of exotic superconducting properties have been predicted in noncentrosymmetric superconductors~\cite{Edelstein1995PRL, Frigeri2004PRL, Fujimoto2007JPSJ, Yanase2007JPSJ}, but many have not yet been observed.



A number of other noncentrosymmetric superconductors have since been reported and are being actively studied, including UIr~\cite{Akazawa2004JPSJ}, CeRhSi$_3$~\cite{Kimura2005PRL}, Ir$_2$Ga$_9$~\cite{Shibayama2007JPSJ}, and B-doped SiC~\cite{Ren2007JPSJ}.  Exotic properties, however, have so far mostly been observed in materials containing cerium, and cerium compounds commonly exhibit unusual magnetism; the magnetism and superconductivity are thought to originate from the same itinerant 4$f$ electrons. Disentangling the not-fully-understood, novel noncentrosymmetric physics from the complications introduced by strongly correlated $f$ electrons of heavy-fermion compounds remains a key challenge, and will require the study of non-magnetic, $f$-electron-free noncentrosymmetric superconductors. Thus far, the only $d$-electron systems in which unconventional behavior has been reported that might arise from the lack of inversion symmetry are Li$_2$Pt$_3$B~\cite{Yuan2006PRL,Nishiyama2007PRL} and LaNiC$_2$~\cite{Hillier2009PRL}.



Recently, nine new noncentrosymmetric superconductors were reported with the same crystal structure as CeRhSi$_3$ (space group $I4mm$)~\cite{Oikawa2008JPSmeeting,Bauer2009PRB}, with chemical formulas $AM$Si$_3$ ($A$ = Ca, Sr, Ba; $M$ = Co, Rh, Ir, Ni, Pd, Pt).  All are thought to be non-magnetic, and none contain active $f$ electrons.  These materials are uniquely valuable because they can be readily compared to Ce-containing superconductors with an identical crystal structure.  Unlike their Ce-based analogs, in which superconductivity only emerges at high pressures once antiferromagnetism has been suppressed~\cite{Kimura2007JPSJ,Sugitani2006JPSJ,Settai2007IJMPB}, the newly-discovered materials superconduct at ambient pressure.  This is suggestive of a magnetic pairing mechanism for the Ce-based materials and a more conventional phonon mechanism for the new materials.


The new $f$-electron-free 1-1-3 silicides have the potential to be the subject of intense research.  A key first step is basic characterization of the superconducting state. To this end, we prepared polycrystalline samples of the Ca$M$Si$_3$ materials ($M$ = Ir, Pt) that would be expected to have the strongest spin-orbit interactions and investigated their structure, specific heat, resistivity, and magnetic properties.


\section{Sample preparation and characterization}
\begin{figure}[b]
\centering
\includegraphics[width=\columnwidth, height=6cm,clip]{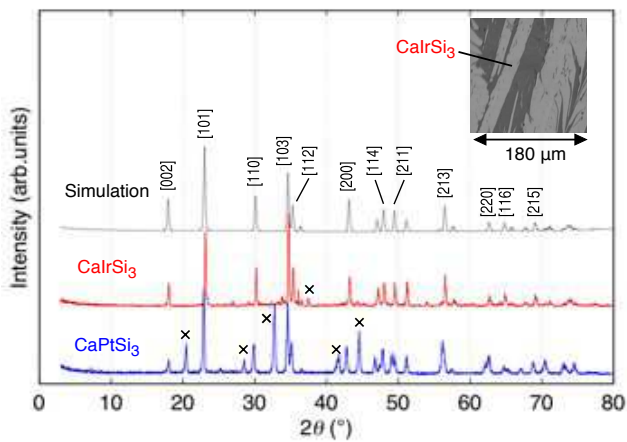}
\caption{(Color online) Powder x-ray diffraction results for CaIrSi$_3$ and CaPtSi$_3$. The crosses indicate impurity peaks. The inset displays a scanning electron micrograph of CaIrSi$_3$. The sample contained an impurity phase. The EDX result reveals the bright area to be CaIrSi$_3$ and the dark area to have the approximate cation ratio of Ca:Ir:Si = 1:0:2.28.}
\label{Fig_XRD}
\end{figure}

Polycrystalline samples were prepared from CaSi (99.9\%, Furuuchi Chemical, 1-3 mm granule), Ir (99.99\%, Furuuchi Chemical, powder), Pt (99.98\%, Nilaco, powder), and Si (99.999\%, Furuuchi Chemical, powder) by arc melting. The raw materials were ground under nitrogen to avoid oxidation, then pressed into a pellet, and melted under argon. Samples synthesized starting from the nominal composition of Ca:Ir:Si=1:1:3 produced the new material CaIr$_3$Si$_7$, discussed in the Appendix. To compensate for loss of volatile calcium silicides, molar ratios of Ca:Ir:Si=3:1:4.7, and Ca:Pt:Si=3:1:5 were used.
Powder x-ray diffraction (XRD) was performed with a commercial x-ray diffractometer (MAC Science, M03XHT$^{22}$) using CuK$\alpha$ radiation (wavelength: 1.54056 \AA), and surface analyses using a commercial scanning electron microscope (SEM, Keyence VE-9800S) and energy-dispersive x-ray spectrometer (EDX, EDAX VE7800), all at room temperature. The results are presented in Fig.~\ref{Fig_XRD}. Both the CaIrSi$_3$ and CaPtSi$_3$ samples exhibited additional XRD peaks from impurity phases, but were of high enough quality that their superconducting properties could be investigated. The inset of Fig.~\ref{Fig_XRD} displays a SEM image. EDX composition mapping identified two distinct phases in the CaIrSi$_3$ sample, with approximate cation ratios Ca:Ir:Si = 1:1.03:2.95, consistent with CaIrSi$_3$, and Ca:Ir:Si = 1:0:2.28. The Ir-free phase did not appear in XRD, suggesting it to be solidified melt without a well-defined crystal structure.

Single phase CaIrSi$_3$ grains  of size $\sim$200~$\rm{\mu}$m were isolated
using   dilute   hydrochloric   acid,   then   crushed.    Homogeneous
micron-sized grains  were selected  using an ethanol  suspension, then
sealed in a Lindemann glass  capillary of internal diameter 0.2~mm. 
High-resolution synchrotron XRD measurements were performed in transmission mode at  the SPring-8 BL02B2 beam line using N$_2$ gas-flow  temperature control and a  large Debye-Scherrer camera
with  an  imaging   plate  detector  \cite{NishiboriNIMPRA2001} which was  set up  to collect data  from 0.010 to  78.68$^\circ$ in
$2\theta$, with a resolution of 0.010$^\circ$.  Data were collected at
a series  of temperatures from  90 to 300~K  with exposure times of five
minutes, and a  higher-statistics dataset with a 60-minute exposure time
was taken  on a  different sample at  100~K for  structure refinement.
The  highest peak in  the latter  dataset had  approximately 1,100,000
counts.  An incident x-ray   wavelength  of  0.35747(1)~\AA\ was  used,
calibrated with a CeO$_2$ standard.

No technique  was found that  would result in  single-phase CaPtSi$_3$
grains,  but aqua  regia dissolved  most impurities,  and it  was also
possible to  isolate CaPtSi$_3$-free grains of  the remaining impurity
phases.  Spectra with and without CaPtSi$_3$ were collected  as for CaIrSi$_3$, using a wavelength of  0.351190(12)~\AA,  but  without  a  high-statistics  dataset.
By comparing  the CaPtSi$_3$ and  impurity-phase spectra, reflections
associated with impurity phases could be deleted, allowing structure refinements for CaPtSi$_3$ as shown in Fig.~\ref{Fig_XRD}. The previously mentioned limitations require that the CaPtSi$_3$ XRD results be treated with caution, but they should provide a useful starting point. 

\begin{figure}[htb]
\begin{center}
\includegraphics[width=\columnwidth,clip]{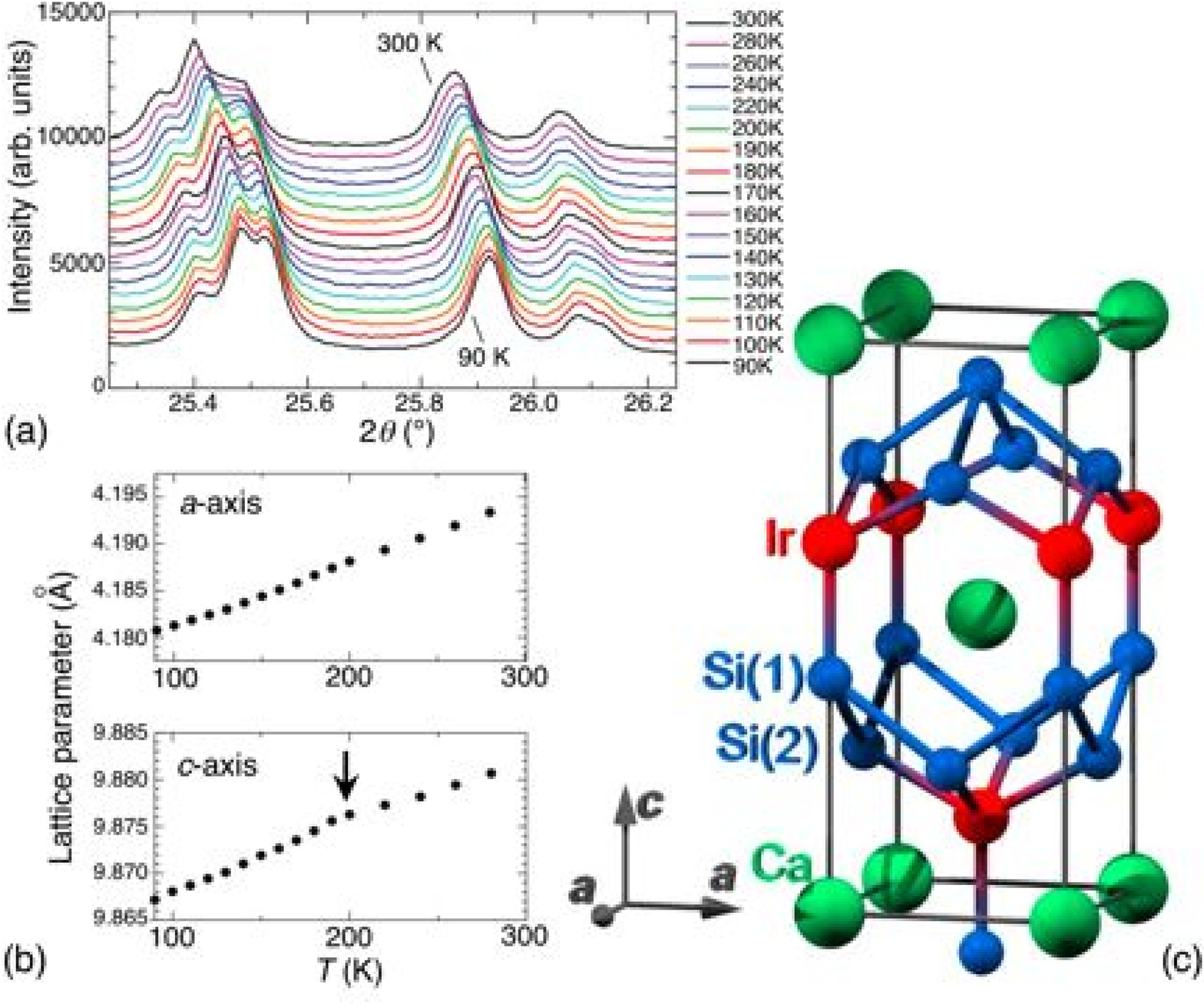}
\caption{(Color online) X-ray diffraction  results  on CaIrSi$_3$: (a)Temperature  dependence of  several powder XRD  peaks of CaIrSi$_3$ from  90 (bottom) to
  300~K (top) with an incident x-ray wavelength of 0.35747(1)~\AA. (b) Temperature dependence  of both lattice parameters. A slight slope change is visible for the $c$~axis at 200~K. (c) Noncentrosymmetric crystal structure of CaIrSi$_3$ as determined by powder x-ray diffraction.}
\label{Fig_XRD_SPring8}
\end{center}
\end{figure}

Both  compounds'  lattice  parameters  increase   monotonically  with
temperature by less  than 0.4\% over the temperature  range probed, as
displayed  in  Fig.~\ref{Fig_XRD_SPring8}  for  CaIrSi$_3$.   In  this
compound only, a small slope  change in the $c$-axis lattice parameter
can be seen  at 200~K, and the $c$-axis position  of the Si(2) (Ir--Si
layer) site appears to fall by about 0.005~\AA\ above this temperature,
making  the Ir--Si  layer slightly  flatter.  The  above  changes were
reflected in the interatomic distances.

\begin{figure}[htb]
\begin{center}
\includegraphics[width=\columnwidth,clip]{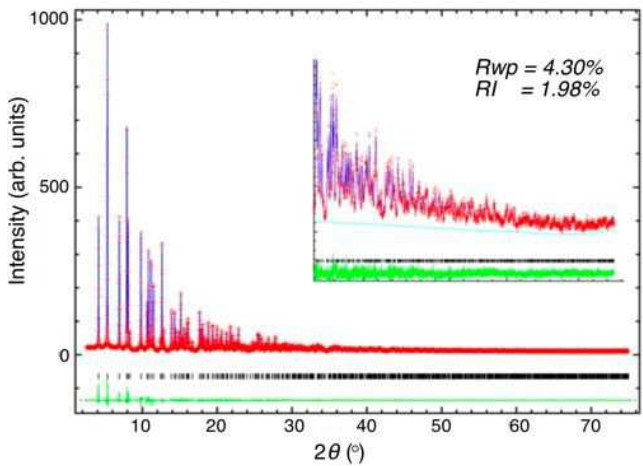}
\caption{(Color online) The Rietveld structure  refinement of CaIrSi$_3$ performed  on  the higher-statistics synchrotron x-ray data at 100~K.}
\label{Fig_Hs_Rietveld}
\end{center}
\end{figure}

Figure \ref{Fig_Hs_Rietveld} and Table~\ref{crystalproperties1} report the results of
a  Rietveld structure  refinement performed  on  the higher-statistics CaIrSi$_3$ 
data  using the program  Synchrotron-Powder \cite{NishiboriActaC2007},
using 1086 reflections from  2.500 to 75.000$^\circ$.  The reliability
factors based on the weighted profile, $R_\text{wp}$, and on the Bragg
intensities, $R_\text{I}$, were 4.30\%\ and 1.98\%, respectively. 
Refinements  of the  temperature-dependent  CaPtSi$_3$ data  typically
produced  reliability   factors  $R_\text{wp}$  and   $R_\text{I}$  of
4.8--5.1\%\ and 7.0--8.5\%, respectively.   The results of the refinement at 100~K are
reported in Table~\ref{crystalpropertiesPt}.


\begin{table}[htb]
\caption{The lattice parameters, fractional atomic coordinates, and interatomic  distances of CaIrSi$_3$ at 100~K.  $U$(iso)  is the isotropic  thermal  displacement parameter.}
\label{crystalproperties1}
\begin{center}
\begin{tabular}{llr@{.}lr@{.}lr@{.}lr@{.}l} \hline
\multicolumn{10}{l}{Refinement} \\
	& \multicolumn{3}{l}{Space group} & \multicolumn{6}{r}{Tetragonal, $I4mm$ (No.\ 107)} \\
	& \multicolumn{5}{l}{$Z$ / Calculated density} & \multicolumn{4}{r}{2 /  6.085 Mg/m$^3$}\\
	& \multicolumn{7}{l}{Absorption coefficient} & 5&732 mm$^{-1}$\\
	& \multicolumn{6}{l}{Data / restraints / parameters} & \multicolumn{3}{r}{1086 / 0 / 16}\\ \cline{1-6}
\multicolumn{10}{l}{Lattice parameters (\AA)} \\
	& $a$ & 4&18327(2) & \multicolumn{6}{r}{} \\
	& $c$ & 9&87278(7) & \multicolumn{6}{r}{} \\ \cline{1-6}
\multicolumn{10}{l}{Fractional Coordinates} \\
	& & \multicolumn{2}{c}{$x$} & \multicolumn{2}{c}{$y$} & \multicolumn{2}{c}{$z$} & \multicolumn{2}{c}{$U$(iso) (\AA$^2$)} \\
	& Ca & 0&00000 & 0&00000 & 0&00000 & 0&00489(24) \\
	& Ir & 0&00000 & 0&00000 & 0&64666(15) & 0&00247(2) \\
	& Si(1) & 0&00000 & 0&00000 & 0&40975(29) & 0&00451(35) \\
	& Si(2) & 0&00000 & 0&50000 & 0&25886(18) & 0&00451(35) \\ \cline{1-6}
\multicolumn{10}{l}{Interatomic distances (\AA)} \\
	& \multicolumn{3}{l}{Ir--Si(1)}    & 2&3396(1) & \multicolumn{4}{r}{} \\
	& \multicolumn{3}{l}{Ir--Si(2)}    & 2&3674(1) & \multicolumn{4}{r}{} \\
	& \multicolumn{3}{l}{Si(1)--Si(2)} & 2&5669(1) & \multicolumn{4}{r}{} \\
	& \multicolumn{3}{l}{Ca--Si(1)}    & 3&0897(1) & \multicolumn{4}{r}{} \\
	& \multicolumn{3}{l}{Ca--Si(2)}    & 3&1687(1) & \multicolumn{4}{r}{} \\
	& \multicolumn{3}{l}{Ca--Ir}       & 3&2931(1) & \multicolumn{4}{r}{} \\ \hline
\end{tabular}
\end{center}
\end{table}

\begin{table}[htb]
\caption{Lattice parameters, fractional atomic coordinates and 
interatomic  distances  for  CaPtSi$_3$  at  100~K;  $U$(iso)  is  the
isotropic thermal displacement parameter.}
\label{crystalpropertiesPt}
\begin{center}
\begin{tabular}{llr@{.}lr@{.}lr@{.}lr@{.}l} \hline
\multicolumn{10}{l}{Refinement} \\
	&                \multicolumn{3}{l}{Space               group}
	&   \multicolumn{6}{r}{Tetragonal,  $I4mm$   (No.\   107)}  \\
	&       \multicolumn{5}{l}{$Z$,       Calculated      density}
	&        \multicolumn{4}{r}{2,        6.133        Mg/m$^3$}\\
	&    \multicolumn{7}{l}{Absorption   coefficient}    &   5&683
	mm$^{-1}$\\   &   \multicolumn{6}{l}{Data   /   restraints   /
	parameters} & \multicolumn{3}{r}{149 / 0 / 5}\\ \cline{1-6}
\multicolumn{10}{l}{Lattice parameters (\AA)} \\
	& $a$ & 4&19880(10) & \multicolumn{6}{r}{} \\
	& $c$ & 9&8111(4) & \multicolumn{6}{r}{} \\ \cline{1-6}
\multicolumn{10}{l}{Fractional Coordinates} \\
	& & \multicolumn{2}{c}{$x$} & \multicolumn{2}{c}{$y$} & \multicolumn{2}{c}{$z$} & \multicolumn{2}{c}{$U$(iso) (\AA$^2$)} \\
	& Ca & 0&00000 & 0&00000 & 0&00000 & 0&0036(10) \\
	& Pt & 0&00000 & 0&00000 & 0&6429(7) & 0&0036(10) \\
	& Si(1) & 0&00000 & 0&00000 & 0&3955(12) & 0&0036(10) \\
	& Si(2) & 0&00000 & 0&50000 & 0&2577(8) & 0&0036(10) \\ \cline{1-6}
\multicolumn{10}{l}{Interatomic distances (\AA)} \\
	& \multicolumn{3}{l}{Pt--Si(1)}    & 2&427(14) & \multicolumn{4}{r}{} \\
	& \multicolumn{3}{l}{Pt--Si(2)}    & 2&383(5) & \multicolumn{4}{r}{} \\
	& \multicolumn{3}{l}{Si(1)--Si(2)} & 2&497(8) & \multicolumn{4}{r}{} \\
	& \multicolumn{3}{l}{Ca--Si(1)}    & 3&141(4) & \multicolumn{4}{r}{} \\
	& \multicolumn{3}{l}{Ca--Si(2)}    & 3&172(6) & \multicolumn{4}{r}{} \\
	& \multicolumn{3}{l}{Ca--Pt}       & 3&283(3) & \multicolumn{4}{r}{} \\ \hline
\end{tabular}
\end{center}
\end{table}

\section{Superconducting properties}
\subsection{Dc susceptibility and resistivity}
\begin{figure}[t]
\centering
\includegraphics[width=\columnwidth,keepaspectratio,clip]{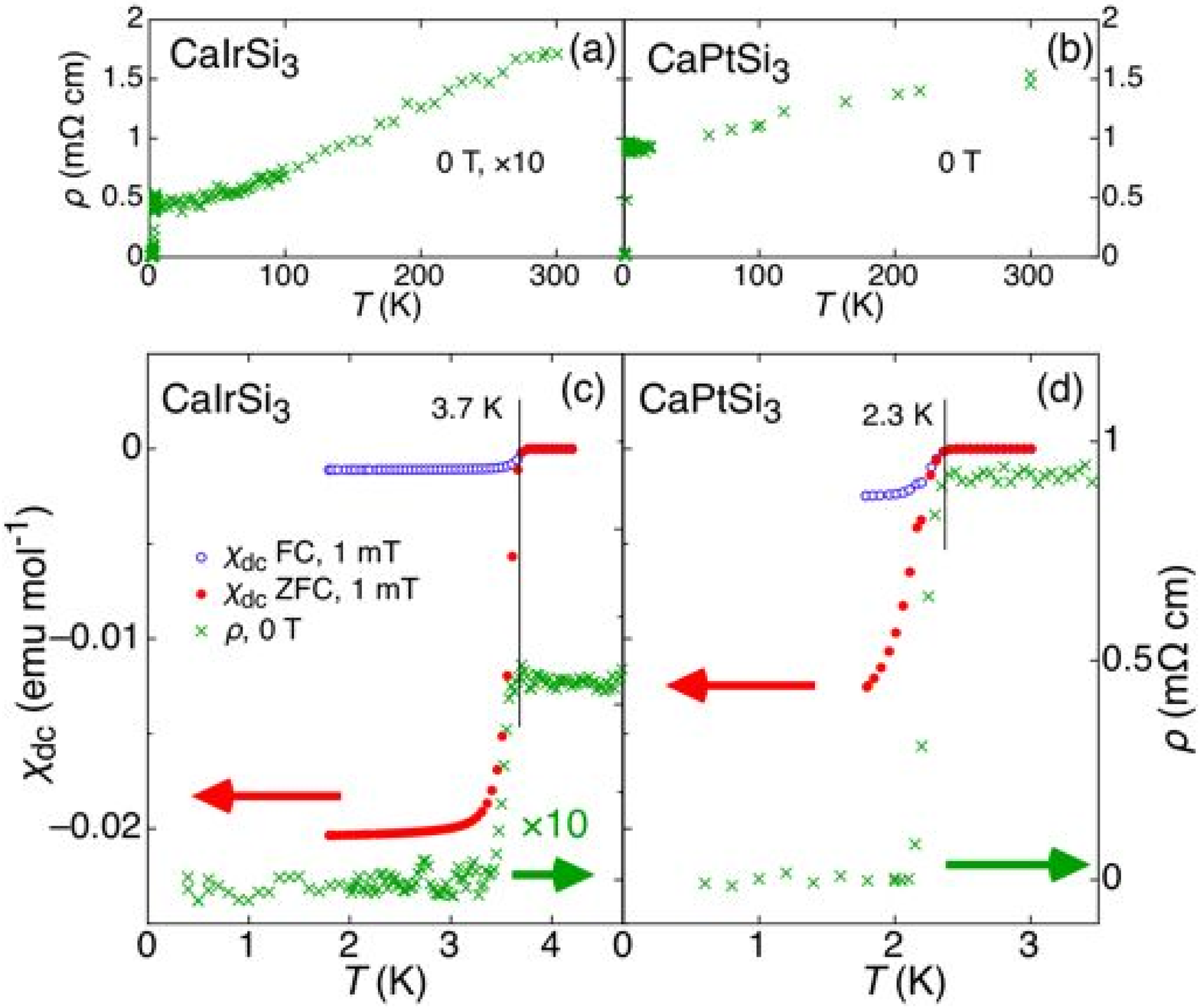}
\caption{(Color online) Temperature dependence of the resistivity $\rho$ for (a) CaIrSi$_3$ and (b) CaPtSi$_3$ up to 300~K and low-temperature dc susceptibility $\chi_{\rm{dc}}$ and resistivity for (c) CaIrSi$_3$ and (d) CaPtSi$_3$. Both zero-field-cooled (ZFC) and field-cooled (FC) dc susceptibility under 1 mT are displayed. The onset temperatures of the superconducting transition are approximately 3.7~K for CaIrSi$_3$ and 2.3~K for CaPtSi$_3$ in both measurements. Strong diamagnetism indicates that the majority of each sample was superconducting.}
\label{Fig_rho_DCchi}
\end{figure}

Using polycrystalline samples, dc susceptibility measurements were performed with a commercial Superconducting Quantum Interference Device (SQUID) magnetometer (Quantum Design, MPMS-XL) down to 1.8~K, and resistivity measurements were performed with a commercial instrument (Quantum Design, PPMS) with a $^3$He refrigeration insert down to 0.35~K using a conventional four-probe technique. Resistivity in zero field up to 300~K for each compound is exhibited in Figs.~\ref{Fig_rho_DCchi} (a) and~\ref{Fig_rho_DCchi} (b). The temperature dependence of the resistivity indicates typical metallic behavior with residual resistivity ratios (RRR, $\rho_{300{\rm{K}}}/\rho_{5{\rm{K}}}$) of $\sim$4 for CaIrSi$_3$ and $\sim$1.6 for CaPtSi$_3$. The low-temperature resistivity and the dc susceptibility in 1 mT for each compound, presented in Figs.~\ref{Fig_rho_DCchi} (c) and~\ref{Fig_rho_DCchi} (d), exhibit clear superconducting transitions at 3.7~K for CaIrSi$_3$ and 2.3~K for CaPtSi$_3$. For CaIrSi$_3$ the transition temperature is approximately 20\% higher than that reported in Ref.~\cite{Eguchi2009PhysC}, attributed to improved sample quality.

The volume fractions estimated from  zero field cooled dc susceptibility data were $\sim$170\% for CaIrSi$_3$ and $\sim$ 100\% for CaPtSi$_3$, assuming the samples to be single phase.
These large volume fractions are attributed to neglecting demagnetizing effects, but suggest that a majority of each sample was superconducting.
The large difference between field-cooled and zero-field-cooled data is likely due to the presence of melt inclusions that can trap magnetic flux.
The RRRs and volume fractions indicate that CaIrSi$_3$ was higher quality than CaPtSi$_3$, consistent with the XRD result.

\subsection{Specific heat}
\begin{figure}[t]
\centering
\includegraphics[width=\columnwidth,keepaspectratio,clip]{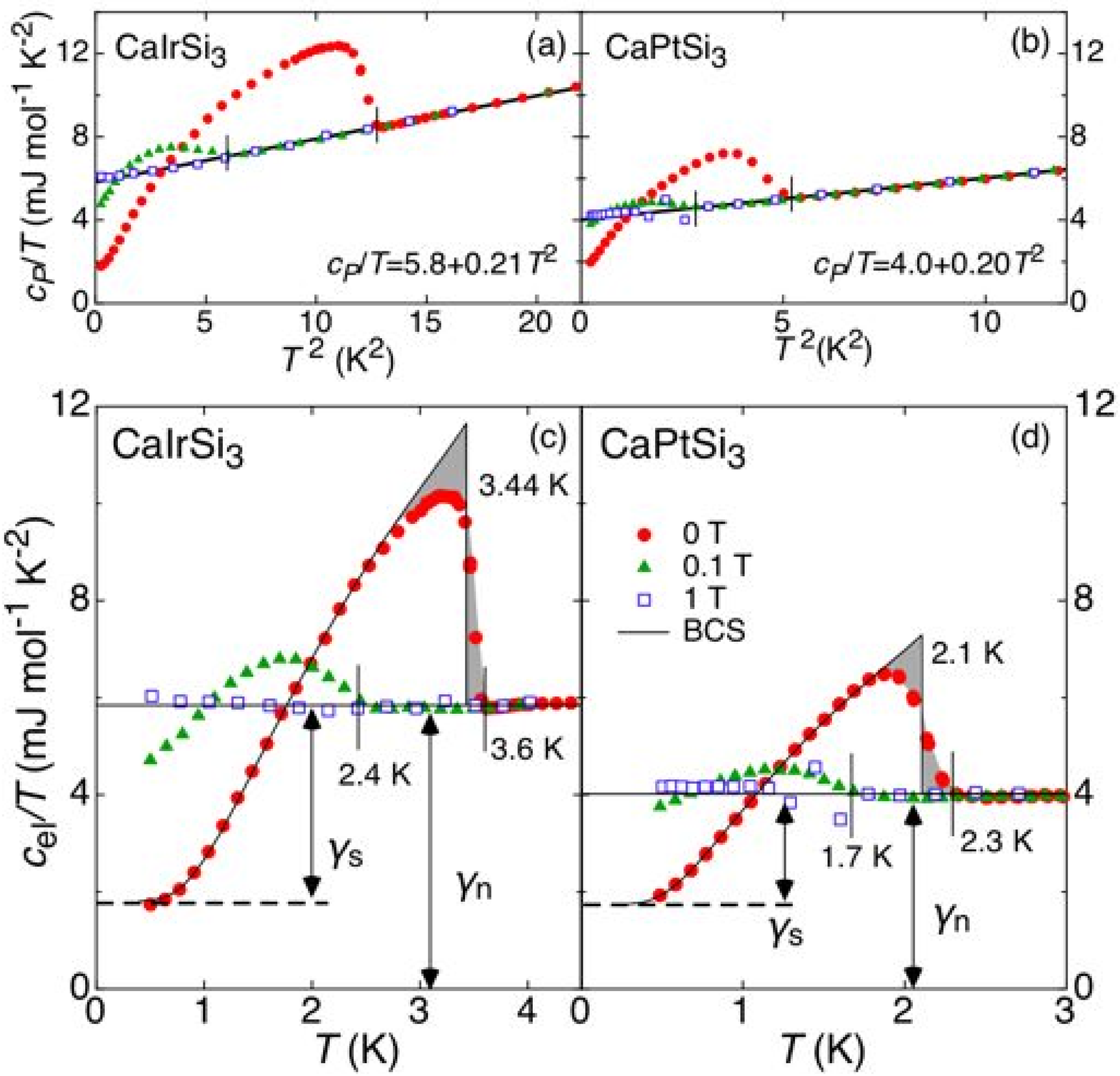}
\caption{(Color online) Temperature dependence of total specific heat $c_P$ for (a) CaIrSi$_3$ and (b) CaPtSi$_3$, and electronic specific heat $c_{\rm{el}}$ for (c) CaIrSi$_3$ and (d) CaPtSi$_3$. $T_{\rm{c}}$ onsets in 0 T are 3.6~K for CaIrSi$_3$ and 2.3~K for CaPtSi$_3$, and in 0.1~T are 2.5~K for CaIrSi$_3$ and 1.7~K for CaPtSi$_3$. Solid lines are BCS fits. The shaded regions have the same area on either side of the specific heat jump, corresponding to entropy conservation. The $T_{\rm{c}}$ values from this procedure match with those from the BCS fits.}
\label{Fig_cel}
\end{figure}

Specific heat measurements were performed down to 0.35~K using a relaxation-time-method calorimeter (Quantum Design, PPMS) on a $^3$He refrigeration insert. The total specific heat for each compound is presented in Figs.~\ref{Fig_cel} (a) and~\ref{Fig_cel} (b), and the electronic specific heat in Figs.~\ref{Fig_cel} (c) and~\ref{Fig_cel} (d). These were calculated assuming the molar weight of the target phases. Superconductivity in both compounds was suppressed by a magnetic field of 1 T. The normal-state specific heat was found to be invariant under external magnetic fields, so the normal-state electronic specific heat coefficients $\gamma_{\rm{n}}$ and the lattice specific heat coefficients $\beta$ were deduced from the data in 1~T by a least-squares fit of the total specific heat $c_P$ to $\gamma_{\rm{n}}T + \beta T^3$. This results in $\gamma_{\rm{n, Ir}}=5.8$ mJ/mol K$^2$ and $\beta_{\rm{Ir}}=0.21$ mJ/mol K$^4$ for CaIrSi$_3$, and $\gamma_{\rm{n, Pt}}=4.0$ mJ/mol K$^2$ and $\beta_{\rm{Pt}}=0.20$ mJ/mol K$^4$ for CaPtSi$_3$. The Debye temperature $\Theta_{\rm{D}}$ of each compound is estimated from $\beta=(12/5)\pi^4N_{\rm{f.u.}}N_{\rm{A}}k_{\rm{B}}/\Theta_{\rm{D}}^3$ as 360~K  for CaIrSi$_3$ and 370~K for CaPtSi$_3$, comparable to those found in ordinary metals. Here $N_{\rm{f.u.}}=5$ is the number of atoms per formula unit, $N_{\rm{A}}$ is Avogadro's number and $k_{\rm{B}}$ is Boltzmann's constant. From the specific heat onsets, we define the transition temperatures $T_{\rm{c}}=3.6$~K for CaIrSi$_3$ and 2.3~K for CaPtSi$_3$ in zero field, consistent with those from dc susceptibility and resistivity. The specific heat jumps indicate the superconductivity to be bulk in nature.

As presented in Figs.~\ref{Fig_cel} (c) and~\ref{Fig_cel} (d), the electronic specific heat coefficient $\gamma$ converges to a finite value at low temperatures even in zero field. This indicates that normal state conduction electrons that do not participate in the superconductivity persist at the Fermi level down to 0~K. However, it cannot be distinguished whether these electrons are contained in CaIrSi$_3$ (CaPtSi$_3$) or in impurity phases.

A fit of $c_{\rm{el}}$ to a polynomial approximation to the conventional weak coupling BCS curve tabulated numerically in Ref.~\cite{Muhlschlegel1959} is also displayed in Figs.~\ref{Fig_cel} (c) and~\ref{Fig_cel} (d). The only two free parameters in this procedure are $T_{\rm{c}}$ and the superconducting contribution to the electronic specific heat coefficient $\gamma_{\rm{s}}$.
The data are fit up to a $T/T_{\rm{c}}$ of just over $70\%$, and agree well with the BCS curve in this region.
The fits result in $T_{\rm{c}}$ values of 3.44~K for CaIrSi$_3$ and 2.1~K for CaPtSi$_3$, and $\gamma_{\rm{s}}$ coefficients of 4.0~mJ/mol K$^2$ for CaIrSi$_3$ and 2.1~mJ/mol K$^2$ for CaPtSi$_3$.
Entropy conservation at the specific heat jump of each compound gives the same $T_{\rm{c}}$ (see Figs.~\ref{Fig_cel} (c) and~\ref{Fig_cel} (d)).
These values of $T_{\rm{c}}$ correspond to an average of the $T_{\rm{c}}$ distribution in each sample.
The superconducting fraction of each sample is estimated from $\gamma_{\rm{s}}/\gamma_{\rm{n}}$ to be $\sim70$\% in CaIrSi$_3$ and $\sim55$\% in CaPtSi$_3$, consistent with earlier indications that the CaPtSi$_3$ sample contained a greater proportion of impurity phase.
The thermodynamic critical field $\mu_{\rm{0}}H_{\rm{c}}(0)$ is estimated using $\mu_0H_{\rm{c}}^2(0)/2=-\gamma_{\rm{s}}T_{\rm{c}}^2/2+\int_0^{T_{\rm{c}}}c_{\rm{el, s}}(T) {\rm{d}}T$, where $c_{\rm{el, s}}(T)$ is the electronic specific heat in the superconducting phase, to be 0.023~T for CaIrSi$_3$ and 0.0094~T for CaPtSi$_3$.

\subsection{Magnetic phase diagram}
\begin{figure}[t]
\centering
\includegraphics[width=\columnwidth,keepaspectratio,clip]{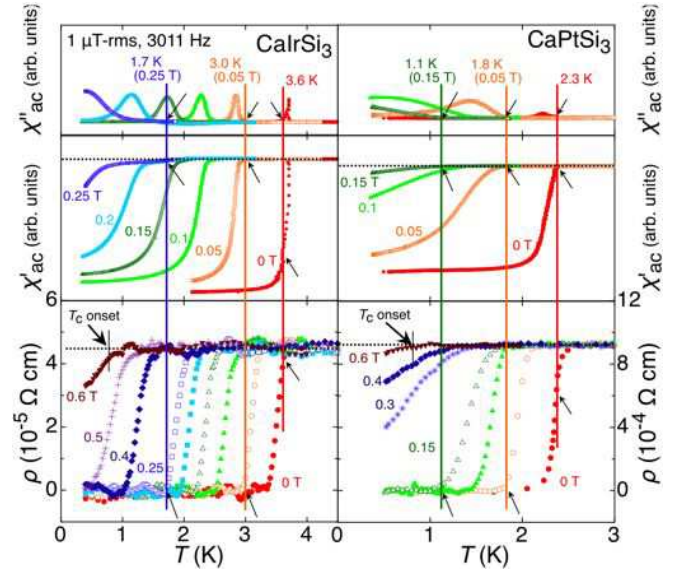}
\caption{(Color online) Temperature dependence of ac susceptibility (the imaginary part $\chi''_{\rm{ac}}$ and the real part $\chi'_{\rm{ac}}$) and resistivity $\rho$ of CaIrSi$_3$ and CaPtSi$_3$ in selected magnetic fields. The zero resistivity temperature matches or exceeds the ac susceptibility onset in magnetic fields. The diamagnetic shielding in $\chi_{\rm{ac}}$ is fully suppressed below 0.3~T for CaIrSi$_3$ and 0.2~T for CaPtSi$_3$, whereas $\rho$ onsets were observed up to 0.6 T.}
\label{Fig_rho_ACchi}
\end{figure}

Measurements of ac susceptibility in various magnetic fields were performed by a mutual-inductance method (1~$\rm{\mu}$T, 3011~Hz) down to 0.3~K using a commercial $^3$He refrigerator (Oxford Instruments, Heliox), equipped with a superconducting magnet. These and the resistivity results are presented in Fig.~\ref{Fig_rho_ACchi}. The ac susceptibility measurements were performed on both warming and cooling with no hysteresis; thus, only the warming results are displayed. Under magnetic fields, ac susceptibility onsets were observed up to 0.27~T for CaIrSi$_3$ and 0.15~T for CaPtSi$_3$, and resistivity onsets up to 0.6~T for both compounds: several times larger than the thermodynamic critical fields. The large critical fields and nonhysteretic transition temperatures clearly indicate that these compounds are type-II superconductors.

\begin{figure}[t]
\centering
\includegraphics[width=6.4cm,keepaspectratio,clip]{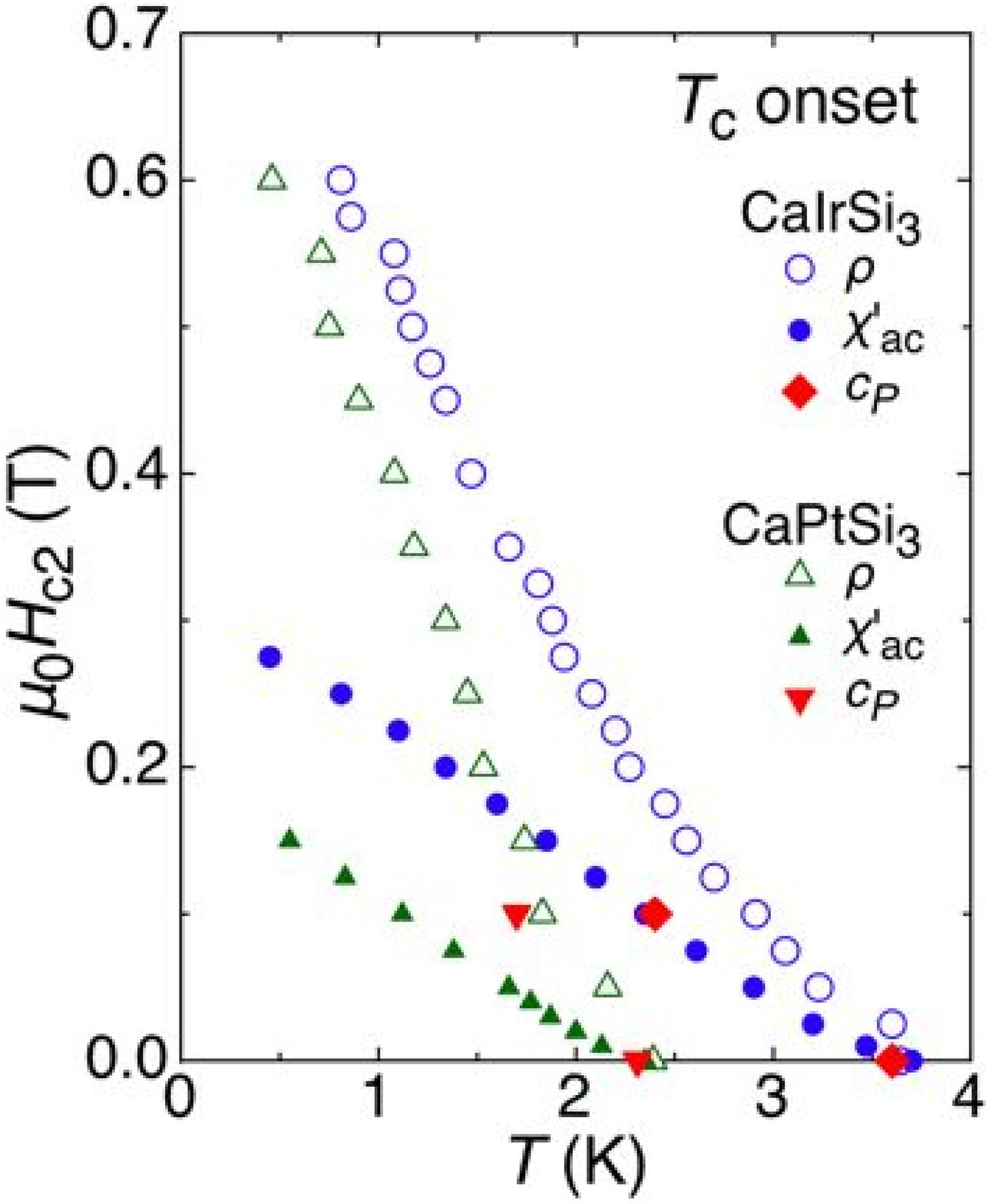}
\caption{(Color online) $H - T$ phase diagram of CaIrSi$_3$ and CaPtSi$_3$. Points represent the onset temperatures in $\chi'_{\rm{ac}}$, $\rho$, and $c_P$, as defined in the text. For both compounds the $H-T$ phase lines as extracted from $\rho$ and $\chi'_{\rm{ac}}$ differ. Possible reasons for this difference are discussed in Sec.~\ref{discussion}.}
\label{Fig_HT}
\end{figure}

The $H - T$ phase diagrams of CaIrSi$_3$ and CaPtSi$_3$ deduced from the ac susceptibility $\chi'_{\rm{ac}}$ onset, resistivity $\rho$ onset, and specific heat $c_P$ onset are summarized in Fig.~\ref{Fig_HT}. The $T_{\rm{c}}$ onsets were defined as 5\% of full, zero-field diamagnetism in $\chi'_{\rm{ac}}$, a 5\% decrease in $\rho$, and the onset in $c_P$. We note that for CaPtSi$_3$ the transition broadens substantially under magnetic fields. This behavior is correlated with the phase purity of the sample as evidenced by impurity peaks in XRD and the larger residual contribution in the specific heat. Each curve is remarkably linear and is slightly concave upward in low magnetic fields. In neither system do the curves deduced from the $\chi'_{\rm{ac}}$ and $\rho$ onsets coincide, and the discrepancy between them is more prominent for CaPtSi$_3$ than for CaIrSi$_3$.

Since the $H - T$ curve of CaIrSi$_3$ based on $\chi'_{\rm{ac}}$ onset begins to saturate at high fields, $\mu_0H_{\rm{c2}}(0)$ is estimated from the points at 0.3 K as a lower limit and the linear extrapolation to zero temperature as an upper limit: 0.27 -- 0.32~T. Similarly for CaPtSi$_3$ it is estimated as 0.15 -- 0.20~T. Approximate Ginzburg-Landau parameters $\kappa_{\rm{GL}}$, superconducting coherence lengths $\xi(0)$, and penetration depths $\lambda(0)$ are estimated from the $\mu_0H_{\rm{c2}}$ lower limits with the relations $\mu_{\rm{0}}H_{\rm{c2}}(0)=\sqrt{2}\kappa_{\rm{GL}}\mu_{\rm{0}}H_{\rm{c}}(0)$, $\mu_0H_{\rm{c2}}(0)=\Phi_{\rm{0}}/2\pi\xi^2(0)$, and $\kappa_{\rm{GL}}=\lambda(0)/\xi(0)$, respectively.
These results should be treated with caution since they come from multiphase, polycrystalline samples of anisotropic materials, but they provide an essential starting point.
The results are summarized in Table~\ref{properties1} with the other physical properties determined in this study.


\begin{table}[t]
\caption{Physical properties of polycrystalline CaIrSi$_3$ and CaPtSi$_3$. Superconducting parameters $\kappa_{\rm{GL}}$, $\xi(0)$, and $\lambda(0)$ are based on the lower limits of $\mu_0H_{\rm{c2}}(0)$ determined from $\chi'_{\rm{ac}}$ onset. Note that these values could be affected by sample quality, and in some cases should be anisotropic.}
\label{properties1}
\begin{center}
	\begin{tabular}{lcc} \hline
			& CaIrSi$_3$ &CaPtSi$_3$ \\ \hline
		$T_{\rm{c}}$ & 3.6 K & 2.3 K \\
\\
		$\mu_0H_{\rm{c2}}(0)$ ($\chi'_{\rm{ac}}$ onset) & 0.27 T & 0.15 T \\
		$\mu_0H_{\rm{c}}(0)$ & 0.023 T & 0.0094 T \\
\\
		$\kappa_{\rm{GL}}$ & 8.3 & 11 \\
		$\xi(0)$& 34 nm & 47 nm  \\
		$\lambda(0)$& 280 nm & 520 nm  \\
\\
		$\gamma_{\rm{n}}$ & 5.8 mJ/mol K$^2$ & 4.0 mJ/mol K$^2$ \\ 
		$\gamma_{\rm{s}}$ & 4.0 mJ/mol K$^2$ & 2.1 mJ/mol K$^2$ \\ 
		$\Theta_{\rm{D}}$ & 360 K & 370 K \\  \hline
	\end{tabular}
\end{center}
\end{table}

\section{Discussion and Conclusion}
\label{discussion}

The specific heat results are generally well explained by weak-coupling BCS theory and similar to reports on BaPtSi$_3$ \cite{Bauer2009PRB}.  While the agreement with the BCS curve suggests that these superconductors are fully gapped, the specific heat of CaIrSi$_3$ falls below this curve 
just above 0.7$T_{\rm c}$. Although the superconducting transition is broad, this is well outside the transition width. This deviation is suggestive of a departure from a pure $s$-wave gap structure, as in a multiband or anisotropic-gap scenario, the alternative being an unusual distribution of $T_{\rm c}$ with a significant tail to low temperatures.
The apparent full gap strongly suggests that the superconducting pairing is dominantly or exclusively singlet; a triplet component could introduce gap anisotropy.

The experimental values of $\mu_0H_{\rm c2}(T)$ are much smaller than the Pauli-limiting fields $\mu_0H_{\rm P}(0)$ of $1.84T_{\rm c}$ expected in weak-coupling $s$-wave BCS theory, $\sim$6.5~T for CaIrSi$_3$ and $\sim$4~T for CaPtSi$_3$. Barring large corrections to $\mu_0H_{\rm P}$, the dominant pair-breaking effect in CaIrSi$_3$ and CaPtSi$_3$ would be orbital depairing.
This represents a significant difference from the Ce-containing isostructural superconductors CeRhSi$_3$, CeIrSi$_3$ and CeCoGe$_3$, in which heavy-fermion masses strongly suppress orbital depairing to reveal that Pauli-limiting behavior does not set in where expected for a singlet condensate.  Pauli pair breaking is strongly suppressed by the presence of a triplet component or a van Vleck-like susceptibility specific to noncentrosymmetric systems~\cite{Fujimoto2007JPSJ, Sugitani2006JPSJ, Settai2007IJMPB, Kimura2007PRL}.
Unless orbital depairing is also suppressed, however, an extremely high $\mu_0H_{\rm c2}(0)$ that violates conventional expectations for the Pauli limit, commonly used as a key signature of novel noncentrosymmetric physics, will not be exhibited. This is the case in CaIrSi$_3$ and CaPtSi$_3$, whose small $\gamma$ values indicate light carrier masses and in which orbital depairing is not suppressed.

In both samples, the $\mu_0H_{\rm c2}(T)$ obtained from the ac susceptibility and resistivity onsets differ substantially.  This discrepancy far exceeds the factor of 1.695 associated with $\mu_0H_{\rm c3}$ surface superconductivity and thin limit physics, so other effects must be involved. One possible scenario is that anisotropic superconducting parameters, expected given the tetragonal crystal structure, lead to an upper critical field with a strong, narrow peak for one field direction, producing very low volume fractions of robust superconductivity.  Another possibility is pressure enhancement of the superconductivity at grain boundaries due to thermal expansion, leading to very thin superconducting surface layers offering a pathway for conductivity but with negligible volume fraction. Finally, the samples could contain networks of more-defected material with almost no volume fraction, in which the mean free path $\ell$ is shorter than the intrinsic $\xi_0$.  In such a region, the effective coherence length $\xi = (1/\xi_0 + 1/\ell)^{-1}$ is limited by $\ell$, increasing $\mu_0H_{\rm c2} \sim 1/\xi^2$.  This would also help to explain a discrepancy with our earlier image furnace-grown CaIrSi$_3$ samples \cite{Eguchi2009PhysC}, which have a significantly lower $T_{\rm c}$ and higher $\mu_0H_{\rm c2}(0)$.
If $\xi$ in the majority of the samples in the present study is dominated by the change in $\xi_0$, which decreases with decreasing $T_{\rm{c}}$, the broadening of the transition in $\chi_{\rm{ac}}$ can also be coherently explained. The data currently available do not permit determining which of these effects contribute or are dominant.  Settling these issues will require single-crystalline samples --- anisotropy in the superconducting parameters would be readily apparent when varying the field angle, pressure effects could be studied, and magnetic and non-magnetic impurities could be doped in to test their effect controllably.

An upward curvature in low fields and striking linearity of $\mu_0H_{\rm c2}(T)$ are observed in both compounds. Such behavior is quite uncommon, but has been observed in multiband superconductors, for example, MgB$_2$~\cite{Karpinski2003SST}. With the full crystal structure determined in CaIrSi$_3$ and CaPtSi$_3$, band-structure calculations may now be performed to predict the shape of the Fermi surface and whether multiband physics is likely to play a role. This is noteworthy since, to the authors' knowledge, atomic positions have never been published for the Ce analog, although they have been for LaIrSi$_3$ and LaRhSi$_3$ \cite{Lejay1984MRB}, nonsuperconducting CePtSi$_3$ \cite{Kawai2007JPSJ}, and BaPtSi$_3$~\cite{Bauer2009PRB}.

One somewhat surprising result is that CaIrSi$_3$, CaPtSi$_3$ and previously published BaPtSi$_3$ \cite{Bauer2009PRB} all behave in a very similar fashion.  The only clear qualitative differences are in the transition widths and the magnitude of the difference between the $T_{\rm c}$ onsets.  Because the spin-orbit coupling strength is expected to increase strongly with atomic number, all three materials are expected to exhibit strong band splitting. Although the Pt-based materials should have somewhat stronger band splitting, its effect on the physical properties will hinge on details of the band structure. In order to determine whether or not noncentrosymmetric physics is operative in these materials, microscopic techniques such as nuclear magnetic resonance or $\mu$SR may prove useful.

In conclusion, we reported the crystallographic and superconducting properties of CaIrSi$_3$ and CaPtSi$_3$. The electronic specific heat coefficients $\gamma$ are a few mJ/mol K$^2$ in both CaIrSi$_3$ and CaPtSi$_3$, values not unusual for metals, indicating that the electron correlations are not strong in these compounds. Their specific heat results suggest that these superconductors are fully gapped. The upper critical fields $\mu_0H_{\rm c2}(0)$ are less than a Tesla and consistent with a conventional orbital depairing mechanism. This and the small $\gamma$ values, constitute a significant departure from the heavy fermion Ce-based materials. Because several results have multiple possible interpretations and the role of anisotropy is unclear, single-crystalline samples will be required. Our results on the promising 5$d$-electron analogs of known Ce-based materials lay the groundwork for studies of the importance of  heavy electrons in noncentrosymmetric superconductors.

\section{Acknowledgements}
We thank S. Yonezawa, H. Takatsu, and S. Kittaka, Y. Tada, S. Fujimoto, C. Michioka, and K. Yoshimura for fruitful discussions and useful advice. The synchrotron radiation experiments performed at BL02B1 and BL02B2 at SPring-8 were supported by the Japan Synchrotron
Radiation Research Institute (JASRI) (Proposal No. 2009A,B0083,0084). This work is supported by a Grant-in-Aid from the Global COE program ``The Next Generation of Physics, Spun from Universality and Emergence'' 
from the Ministry of Education, Culture, Sports, Science, and Technology (MEXT) of Japan, and by the ``Topological Quantum Phenomena" Grant-in Aid for Scientific Research on Innovative Areas from MEXT of Japan.
G. E., M. K., and D. C. P. are supported by the Japan Society for the Promotion of Science (JSPS).

\appendix
\section{The new material CaIr$_3$Si$_7$}

While  optimizing  the   preparation  technique  for  CaIrSi$_3$,  the
previously  unreported ternary  phase  CaIr$_3$Si$_7$ was  discovered, and subsequently reproduced by solid-state reaction of Ca:Ir:Si~=~1:1:3 under vacuum in a sealed quartz tube (1100~$^\circ$C, 24 hours).
This compound crystallizes in  the rhombohedral space group $R\bar 3c$
(No.\  167),    isotypic   to   ScRh$_3$Si$_7$    and   ScIr$_3$Si$_7$
\cite{Chabot1981}, and possesses an inversion center.  It does not exhibit superconductivity above 1.8~K.
We note that thorough investigation of the synthesis of CaPtSi$_3$ by a flux method by Takeuchi \textit{et al.}, resulted in synthesizing a new centrosymmetric superconductor, Ca$_2$Pt$_3$Si$_5$~\cite{Takeuchi2009JPSJ}.

A single-crystalline  grain of  CaIr$_3$Si$_7$, prepared by  melting a
mixture  of stoichiometry  Ca:Ir:Si~=~1:1:3 in  an image  furnace, was
separated  from excess  melt  phases using  dilute hydrochloric  acid.
This  grain  was then  characterized  at  room  temperature by  single-crystal   x-ray    diffraction   at   SPring-8    beam   line   BL02B1
\cite{Noda1998JSR} using  a large  cylindrical  image plate
detector.   A  wavelength of  0.35100~\AA\  was  used,  and data  were
collected over the index range $0\leq h\leq 16$, $0\leq k\leq 16$, and
$0\leq  l\leq  49$,   and  over  a  range  in   $\theta$  of  1.83  to
26.02$^\circ$.   A total of  1816 reflections  were collected,  all of
which were unique and 1697  of which had intensity $>2\sigma(I)$.  The
structure  was  solved  and   refined  using  SHELXS97  and  SHELXL97,
respectively   \cite{Sheldrick2008};   results   are   summarized   in
Table~\ref{crystalproperties3}.   Refinement   was  by  a  full-matrix
least-squares  minimization of $F^2$.   The final  reliability factors
$R_1$ and $wR_2$ were 2.04\%\  and 7.11\%, respectively, on peaks with
intensity   greater  than  $2\sigma(I)$,   and  2.70\%\   and  7.93\%,
respectively, on all data.

\begin{figure}[htb]
\includegraphics[width=5.5cm,clip]{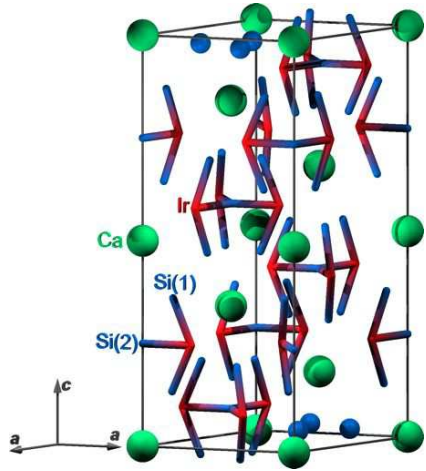}
\caption{(Color online) \label{fig:137structure}Crystal  structure  of CaIr$_3$Si$_7$
  as determined by single-crystal x-ray diffraction.}
\end{figure}

\begin{table}[htb]
\caption{\label{crystalproperties3}Crystal    data    and    structure
  refinement parameters of CaIr$_3$Si$_7$ at 100~K. The  equivalent isotropic
  displacement parameter $U$(eq) is defined  as one third of the trace
  of the orthogonalized $U_{ij}$ tensor.}
\begin{center}
\begin{tabular}{llr@{.}lr@{.}lr@{.}lr@{.}l} \hline
\multicolumn{10}{l}{Refinement} \\
	& \multicolumn{3}{l}{Space group} & \multicolumn{6}{r}{Rhombohedral, $R\bar3c$ (No.\ 167)} \\
	& \multicolumn{5}{l}{$Z$ / Calculated density} & \multicolumn{4}{r}{6 /  8.143 Mg/m$^3$}\\
	& \multicolumn{6}{l}{Absorption coefficient} & \multicolumn{2}{r@{.}}{33}&609 mm$^{-1}$\\
	& \multicolumn{5}{l}{Data / restraints / parameters} & \multicolumn{4}{r}{1816 / 0 / 20}\\
	& \multicolumn{7}{l}{Extinction coefficient} & 0&00016(15)\\ \cline{1-6}
\multicolumn{10}{l}{Lattice parameters (\AA)} \\
	& $a$ & 7&5782(3) & \multicolumn{6}{r}{} \\
	& $c$ & 20&0091(2) & \multicolumn{6}{r}{} \\ \cline{1-6}
\multicolumn{10}{l}{Fractional Coordinates} \\
	& & \multicolumn{2}{c}{$x$} & \multicolumn{2}{c}{$y$} & \multicolumn{2}{c}{$z$} & \multicolumn{2}{c}{$U$(eq) (\AA$^2$)} \\
	& Ir    & 0&346499(17) & 0&013165(17) & 0&08333     & 0&00205(3)\\
	& Ca    & 0&00000      & 0&00000      & 0&00000     & 0&00308(10)\\
	& Si(1) & 0&46234(11)  & 0&14240(11)  & -0&02962(4) & 0&00338(8)\\
	& Si(2) & 0&66667      & 0&33333      & 0&08333     & 0&00237(17)\\ \cline{1-6}
\multicolumn{10}{l}{Interatomic distances (\AA)} \\
 & \multicolumn{3}{l}{Ir--Si(2)}    & 2&42630(13) & \multicolumn{4}{r}{} \\
 & \multicolumn{3}{l}{Ir--Si(1)}    & 2&4449(7) & \multicolumn{4}{r}{} \\
 & \multicolumn{3}{l}{Ir--Si(1)}    & 2&4526(7) & \multicolumn{4}{r}{} \\
 & \multicolumn{3}{l}{Ir--Si(1)}    & 2&5201(7) & \multicolumn{4}{r}{} \\
 & \multicolumn{3}{l}{Ir--Ca}       & 3&0697(1) & \multicolumn{4}{r}{} \\
 & \multicolumn{3}{l}{Ca--Si(1)}    & 3&1257(7) & \multicolumn{4}{r}{} \\
 & \multicolumn{3}{l}{Si(1)--Si(1)} & 2&5985(13) & \multicolumn{4}{r}{} \\
 & \multicolumn{3}{l}{Si(1)--Si(2)} & 2&7127(7) & \multicolumn{4}{r}{} \\
 & \multicolumn{3}{l}{Si(1)--Si(1)} & 2&7365(15) & \multicolumn{4}{r}{} \\
 & \multicolumn{3}{l}{Si(1)--Si(1)} & 2&7605(15) & \multicolumn{4}{r}{} \\ \hline
\end{tabular}
\end{center}
\end{table}

Since this crystal structure contains a large number of atoms per unit
cell in a  nontrivial arrangement, it must be  broken into subunits to
be  described.  As  ScRh$_3$Si$_7$, it  was portrayed  as an  array of
Sc(Ca)-centered Rh(Ir) octahedra and Si double tetrahedra in which the
shared apex was the  Si(2) site \cite{Chabot1981}.  However, since the
shortest bonds in the material are those between Rh(Ir) and Si, it may
be  more realistic  to visualize  the structure  in terms  of slightly
twisted  Si-centered Ir$_3$Si$_7$  barrels and  isolated Ca  atoms.  A
crystal   structure   based   on   this   scheme   is   presented   in
Fig.~\ref{fig:137structure}.

\bibliographystyle{apsrev4-1_nocomma}
\bibliography{Preload,myrefs}

\end{document}